# Evaluation of the acoustic and non-acoustic properties of sound absorbing materials using a three-microphone impedance tube

Olivier Doutres, Yacoubou Salissou, Noureddine Atalla, Raymond Panneton

GAUS, Department of mechanical engineering, Université de Sherbrooke (Qc), Canada, J1K 2R1

Email address: olivier.doutres@usherbrooke.ca

**ABSTRACT**

This paper presents a straightforward application of an indirect method based on a three-microphone impedance tube setup to determine the non-acoustic properties of a sound absorbing porous material. First, a three-microphone impedance tube technique is used to measure some acoustic properties of the material (i.e., sound absorption coefficient, sound transmission loss, effective density and effective bulk modulus) regarded here as an equivalent fluid. Second, an indirect characterization allows one to extract its non-acoustic properties (i.e., static airflow resistivity, tortuosity, viscous and thermal characteristic lengths) from the measured effective properties and the material open porosity. The procedure is applied to four different sound absorbing materials and results of the characterization are compared with existing direct and inverse methods. Predictions of the acoustic behavior using an equivalent fluid model and the found non-acoustic properties are in good agreement with impedance tube measurements.

**1. Introduction**

Characterization of sound absorbing materials, like mineral wool or polymer foam, in the context of building or transport applications, requires the evaluation of their acoustic and non-acoustic (or macroscopic) properties. The acoustic properties evaluate the material sound absorbing efficiency, whereas the non-acoustic properties allow one to predict the material acoustic response in various industrial applications by the use of an appropriate model. In this paper, an equivalent fluid model [1] is used to describe the sound propagation in a rigid or limp frame porous with the associated non-acoustic properties: static airflow resistivity $\sigma$, porosity $\phi$, tortuosity $\alpha_\infty$, viscous characteristic length $\Lambda$, and thermal characteristic length $\Lambda'$.

Classical methods to evaluate non-acoustic properties of porous materials can be sorted in three groups: (1) the direct methods based on the physical definition of the searched property (see examples for $\phi$, $\sigma$, $\alpha_\infty$ and $\Lambda'$ in references [2-5], respectively); (2) the indirect methods based on the acoustical model from which analytical expressions linking the material non-acoustic properties to acoustical measurements are derived [6-9]; and (3) the inverse methods based on an optimization problem where the properties are adjusted in the model to reproduce acoustic measurements [10,11]. In the case of the direct methods, measurement of all non-acoustical properties is not straightforward because one dedicated setup per property is required. The two other types of methods are based on impedance tube or ultrasound measurements. In this paper, only the indirect [6,7] and inverse [10] methods based on impedance tube measurements will be addressed. While the inverse method generally uses a surface acoustic property to operate (e.g., sound absorption coefficient or surface impedance), the indirect method needs two intrinsic acoustic properties of the material, such as the effective density $\rho$ and the effective bulk modulus $K$, usually obtained with an impedance tube setup that can be relatively heavy (e.g., use of an anechoic termination or two different terminations, up to four microphones and six transfer function measurements).

This paper proposes and tests a straightforward procedure for the application of the acoustical indirect method to evaluate the non-acoustic properties of a sound absorbing material by the use of a recently proposed three-microphone impedance tube method [12]. This three-microphone method was shown to be less heavy and more accurate than other existing methods to measure the effective acoustic properties. The proposed straightforward procedure involves: (i) the direct measurement of the open porosity; (ii) the measurement of the acoustic effective properties using the three-microphone impedance tube in the frequency bands where the material behaves as an equivalent fluid; (iii) and the evaluation of the macroscopic non-acoustic properties using the indirect method. This straightforward procedure is applied to four sound absorbing materials frequently used in the context of transport or building applications. These materials have been selected because of their distinct acoustic behavior related to their porous microstructure, i.e. two materials are foams constituted of a continuous arrangement of cells reticulated or not, and the two other material are fibrous constituted of a discontinuous stack of fibers.

The first part of the paper describes the principle of measurement. The experimental setup and the porous materials used for the characterization method are then presented. Results evaluated with the indirect method for the four tested materials are finally compared with those given by existing direct and inverse methods.

## 2. Principle of measurement

### 2.1 Determination of the acoustic properties

The three-microphone method proposed by Salissou [12] allows one to simultaneously determine the normal incidence sound absorption coefficient $\alpha$, the normal incidence sound transmission loss coefficient $nSTL$, and the effective acoustic properties of the tested porous material by the impedance tube setup shown in Fig. 1. In this configuration, the porous sample is backed on the rigid termination. Here the sample is assumed to be homogeneous, symmetric, isotropic and acoustically rigid or limp (i.e., it behaves as an equivalent fluid [1]).

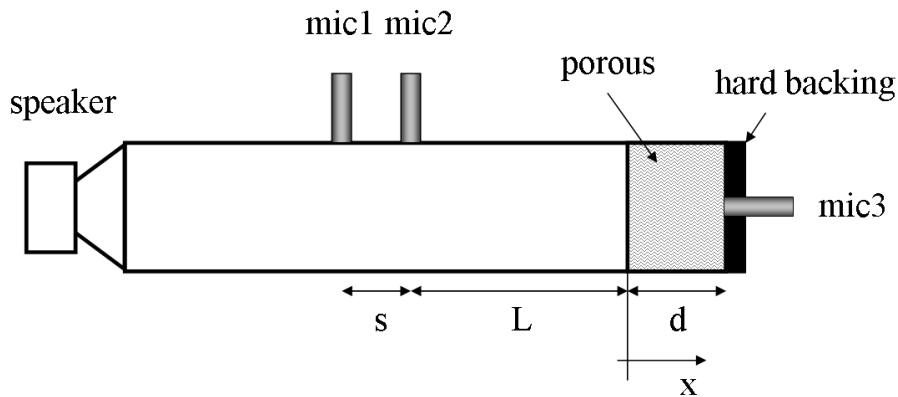

Figure 1. *Standing wave tube with 3 microphones.*

From the two pressure transfer function measurements $H_{12}$ and $H_{23}$, respectively between microphones 2 and 1 and microphones 3 and 2, one can deduce the pressure ratio between the front ($x=0$) and the rear face ($x=d$) of the porous layer as

$$H_{0d} = \frac{1+R_x}{e^{jk_0 L} + R_x e^{-jk_0 L}} H_{23} \tag{1}$$

with $R_x$ the complex reflection coefficient given by

$$R_x = \frac{e^{jk_0 s} - H_{12}}{H_{12} - e^{-jk_0 s}} e^{2jk_0 L}$$

(2)

Here, $k_0$ represents the wave number in the ambient fluid, $s$ is the spacing between microphones 1 and 2, $L$ is the distance between microphone 2 and the front surface of the porous sample and $d$ is the thickness of the sample.

From the transfer matrix approach [13] and considering that the velocity of the air particle at $x = d$ is equal to zero, it is shown that the transfer function $H_{0d}$ is equivalent to the first element of the normal incidence transfer matrix $T_{11}$ in the case of a finite depth layer of homogeneous, isotropic and symmetric porous material. Thus, the wave number and the characteristic impedance of the material can be evaluated as

$$k = \frac{1}{d}\cos^{-1}(H_{d0}) \quad \text{and} \quad Z_c = jZ_s \tan(kd) \ .$$

(3)

The effective density and the effective bulk modulus of the porous material required for the indirect method are thus given by

$$\rho = \phi Z_c k / \omega \quad \text{and} \quad K = \phi \omega Z_c / k$$

(4)

with $\omega$ the angular frequency.

The normal incidence absorption coefficient $\alpha$ of the porous layer is derived from the complex reflection coefficient $R_x$ using:

$$\alpha = 1 - |R_x|^2 \ .$$

(5)

The normal incidence transmission coefficient $\tau_\infty$ is determined from the wave number and the characteristic impedance of the acoustical material as

$$\tau_\infty = 2j \frac{Z_c}{Z_0 - 2jZ_c \cot(kd) + Z_c^2/Z_0} \frac{1}{\sin(kd)} e^{jk_0 d}, \qquad (6)$$

where $Z_0$ is the characteristic impedance of ambient air. Finally, the normal incidence sound transmission loss is obtained from $nSTL = -20\log(|\tau_\infty|)$.

*2.2 Determination of the non-acoustic properties*

The main five non-acoustic properties required in most recent equivalent fluid models [1] for porous materials are the static airflow resistivity $\sigma$, the tortuosity $\alpha_\infty$, the viscous characteristic lengths $\Lambda$, and the thermal characteristic lengths $\Lambda'$. Other properties exist but are not investigated in this paper since the main five properties are usually sufficient for engineering analysis in the context of building or transport applications. Here, these five macroscopic properties are determined using the indirect method proposed by Panneton and Olny [6,7]. In this method, analytical solutions are derived from the Johnson *et al.* viscous model [14] and the Lafarge *et al.* thermal model [15] to extract the macroscopic properties from the measured effective density $\rho$ and bulk modulus $K$. Also, to use the analytical solutions, the open porosity $\phi$ is assumed to be known from a direct measurement [2]. For the sake of completeness, the analytical solutions used in this paper to extract the macroscopic properties are now recalled [6,7]:

$$\sigma = -\frac{1}{\phi} \lim_{\omega \to 0} \left[ \text{Im}(\omega \rho) \right] \tag{7}$$

$$\alpha_\infty = \frac{1}{\rho_0} \left( \text{Re}(\rho) - \sqrt{\text{Im}(\rho)^2 - \left(\frac{\sigma\phi}{\omega}\right)^2} \right) \tag{8}$$

$$\Lambda = \alpha_\infty \sqrt{\frac{2\rho_0 \eta}{\omega \, \text{Im}(\rho)(\rho_0 \alpha_\infty - \text{Re}(\rho))}} \tag{9}$$

$$\Lambda' = \delta_t \sqrt{2} \left( -\text{Im} \left( \left( \frac{\phi - K/K_a}{\phi - \gamma K/K_a} \right)^2 \right) \right)^{-1/2} \tag{10}$$

where $\rho_0$, $K_a$ and $\eta$ represent the density, adiabatic bulk modulus, and dynamic viscosity of the ambient fluid, Pr is the Prandtl number, and $\delta_t = \sqrt{2\eta/\rho_0 \omega \text{Pr}}$. It is worth mentioning that extraction of $\alpha_\infty$, $\Lambda$ and $\Lambda'$ is straightforward, whereas extraction of $\sigma$ is based on a low-frequency extrapolation as explained in the reference paper [6].

The results of the indirect characterization will be compared with existing direct methods and with the inverse acoustical characterization technique [10]. The later method is based on an optimization problem where unknown parameters are adjusted to fit impedance tube measurements of a surface acoustic property; here the normal incidence sound absorption coefficient is used. The method is applied to evaluate the tortuosity and the two characteristic lengths by assuming open porosity and airflow resistivity known from direct measurements.

### 3. Measurement setup

Measurements of the acoustic effective properties ($\rho$, $K$) according to the three-microphone method described in section 2 were carried out in a 44.5-mm diameter impedance tube. A loudspeaker at one end of the tube generated a broadband random signal from 100 Hz to 4200 Hz. The samples were placed at the end of the standing wave tube on the hard termination. Three BSWA Type MPA416 microphones were used as shown in Fig.1. Microphones 1 and 2 were at standard positions with $L = 45$ mm and $s = 25$ mm, whereas microphone 3 is flush mounted on the hard termination. Transfer function $H_{12}$ between microphones 2 and 1 and transfer function $H_{13}$ between microphones 3 and 1, were estimated following the approach described in standard ISO-10534-2 [16]. The transfer function $H_{23}$ was then obtained by the ratio $H_{13}/H_{12}$. To minimize the effects of microphone phase mismatch, a microphone switching calibration procedure was used based on that suggested in the ISO-10534-2 [16].

The direct measurement of the open porosity $\phi$ was performed using the pressure/mass method [2]. From this value and the measurement of the effective properties ($\rho$, $K$), the non-acoustic properties are thus evaluated from the indirect method (i.e., Eqs. (7)-(10)). The static airflow resistivity was determined using the extrapolation method in the low frequency range [6], whereas the tortuosity and the two characteristic lengths were evaluated in the high frequency range by deriving mean values in a specific frequency band where the parameters were relatively constants. Indeed, as mentioned by Panneton and Olny [6], the constancy of the determined parameters in a given frequency range assesses the validity of the used equivalent fluid model in this range. Furthermore, working at low and high frequencies allows one to avoid the influence of the frame vibration which generally occurs at medium frequencies.

For comparison purposes, the static airflow resistivity $\sigma$, was measured using a direct method in accordance to the work by Stinson and Daigle [3] and the tortuosity $\alpha_\infty$ from the ultrasound technique worked out by Allard *et al.* [8]. In the case of the inverse method [10], the measurement of the absorption coefficient was carried out using microphones 1 and 2 according to the standard ISO-10534-2 [16]. Note that frequency bands where frame resonance occurred were also rejected from the inverse characterization process.

Four materials with different pore geometries were investigated. Material A and B were low and high static airflow resistivity plastic foams, respectively, both with a stiff and low density skeleton. Material C and D were low and high density fibrous materials, respectively, both with a soft skeleton and a low static airflow resistivity. These four materials are frequently used in aerospace and building applications for thermal and sound insulation. Some properties of these four materials are listed in Table I; porosity being determined by direct method. Note that all material samples were cut to fit snugly inside the sample holder. However, no additional elements such as sticking nails [7] were used to suppress or minimize the resonant vibrations of the frame. Thus, the frequency bands where the frame had a significant influence on the material acoustic behavior were rejected during the characterization process. Therefore, the constancy of the parameters for the indirect characterization method was obtained between 3 and 4.2 kHz for material A, between 3.8 and 4.2 kHz for material B, and between 1.5 kHz and 3 kHz for materials C and D.

*Table I. Properties of the material samples*

| Material | Thickness (mm) | Porosity $\phi$ | Frame density (kg.m$^{-3}$) |
|---|---|---|---|
| A | 51.44 ± 0.05 | 0.98 ± 0.03 | 9 |
| B | 49.11 ± 0.05 | 0.99 ± 0.03 | 5 |
| C | 18.5 ± 1 | 0.99 ± 0.03 | 5.5 |
| D | 81 ± 1 | 0.99 ± 0.03 | 40 |

## 4. Results and discussion

Table II presents a comparison of the results obtained from the different methods. It is shown that the indirect determination of the static airflow resistivity $\sigma$ is in good agreement with the direct measurements for all materials. The estimation of the tortuosity $\alpha_\infty$ using the indirect method was below the unit value for materials A, C and D. However, since this parameter converged slowly to 1 with increasing the frequency; it was set to unity for these three materials [6]. Note that these results agree with the direct and inverse characterizations and are typical for this kind of materials ($\alpha_\infty \approx 1$ for fibrous [17]). In the case of material B, the characterization of non-acoustic parameters by inverse and indirect methods was difficult since the effect of frame vibration is important in a broad frequency range: between 400 and 3800 Hz. Thus, the inverse method was applied from 200 to 400 Hz and the indirect method (evaluation of $\alpha_\infty$, $\Lambda$ and $\Lambda'$) from 3800 to 4200 Hz; this can explain the relative large difference in the evaluation of the tortuosity $\alpha_\infty$. Furthermore, because of its high airflow resistivity, thin samples of material B were cut to apply the ultrasonic method. Result given by this method validates the use of the indirect method in the high frequency range (above the frame vibration influence). Similar results of the viscous characteristic length $\Lambda$ were also derived from the indirect and inverse methods for all materials: a maximum difference of 15% is found in the case of material D. The determination of the thermal characteristic length $\Lambda'$ using the indirect method is in good agreement with the inverse method for materials A, C and D. For material B, as stated previously, the difference can be attributed to the important influence of the frame on its acoustic behavior. For this particular case, sticking nails could improve the characterization by minimizing the frame vibration [7].

*Table 2. Characterization of the macroscopic properties with the three different techniques*

| Material | Method | Airflow resistivity $\sigma(N.s.m^{-4})$ | Tortuosity $\alpha_\infty$ | Viscous characteristic length $\Lambda$ ($\mu m$) | Thermal characteristic length $\Lambda'$ ($\mu m$) |
|---|---|---|---|---|---|
| A | Direct [a] | 10800 ± 132 | 1.04 ± 0.01 | - | - |
|   | Inverse [b] | - | 1.03 ± 0.04 | 129.0 ± 25.0 | 198.0 ± 94.0 |
|   | Indirect | 10254 ± 434 | 1 | 127.4 ± 4.3 | 185.8 ± 17.1 |
| B | Direct [a] | 44195 ± 1612 | 1.64 ± 0.31 | - | - |
|   | Inverse [b] | - | 1.00 ± 0.01 | 26.3 ± 19.2 | 267.7 ± 99.6 |
|   | Indirect | 39702 ± 3051 | 2.02 ± 0.56 | 26.0 ± 4.3 | 165.8 ± 53.3 |
| C | Direct [a] | 14557 ± 2274 | 1.02 ± 0.03 | - | - |
|   | Inverse [b] | - | 1.03 ± 0.01 | 61.8 ± 6.5 | 110.4 ± 32.2 |
|   | Indirect | 14620 ± 2870 | 1 | 58.4 ± 4.0 | 96.0 ± 15.0 |
| D | Direct [a] | 13430 ± 1744 | 1.04 ± 0.00 | - | - |
|   | Inverse [b] | - | 1.04 ± 0.11 | 54.9 ± 16.9 | 238.7 ± 154.4 |
|   | Indirect | 14379 ± 731 | 1 | 64.5 ± 7.1 | 279.4 ± 33.8 |

[a] For the direct method, the characteristic lengths were not measured.
[b] For the inverse method, the open porosity and the static airflow resistivity were fixed.

Now, let us compare to measurements the normal incidence sound absorption coefficient and sound transmission loss predicted by the five-parameter Johnson-Champoux-Allard equivalent fluid model for materials A and C (see Fig. 2 and 3). Details on this model are reviewed elsewhere [1,6,7]. Predictions are carried out using the non-acoustic properties evaluated from the inverse and indirect methods. It is shown that the predictions based on the two sets of non-acoustic properties give similar results and are in good agreement with the normal incidence measurements. However, the resonant behavior is not predicted due to the rigid frame assumption.

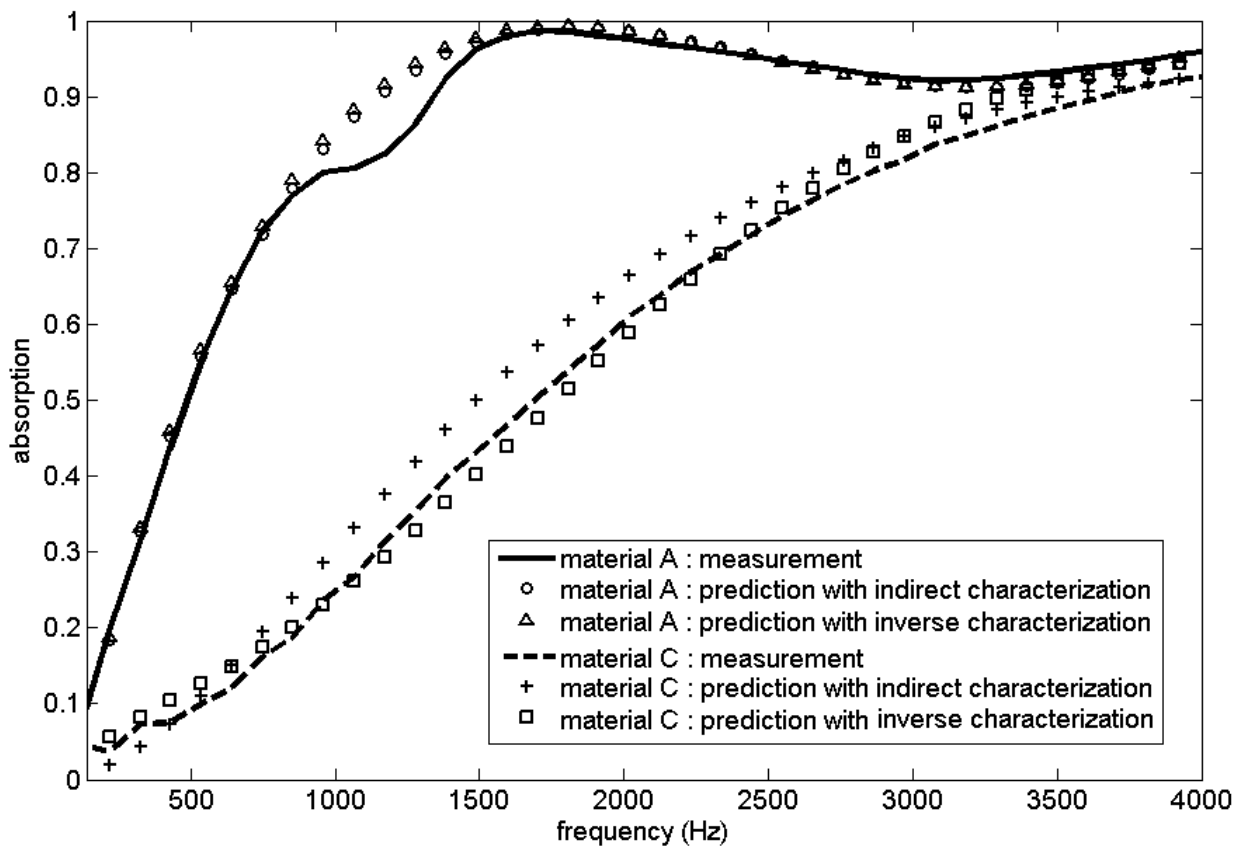

*Figure 2. Normal incidence sound absorption coefficient: comparison between measurements and predictions for materials A and C.*

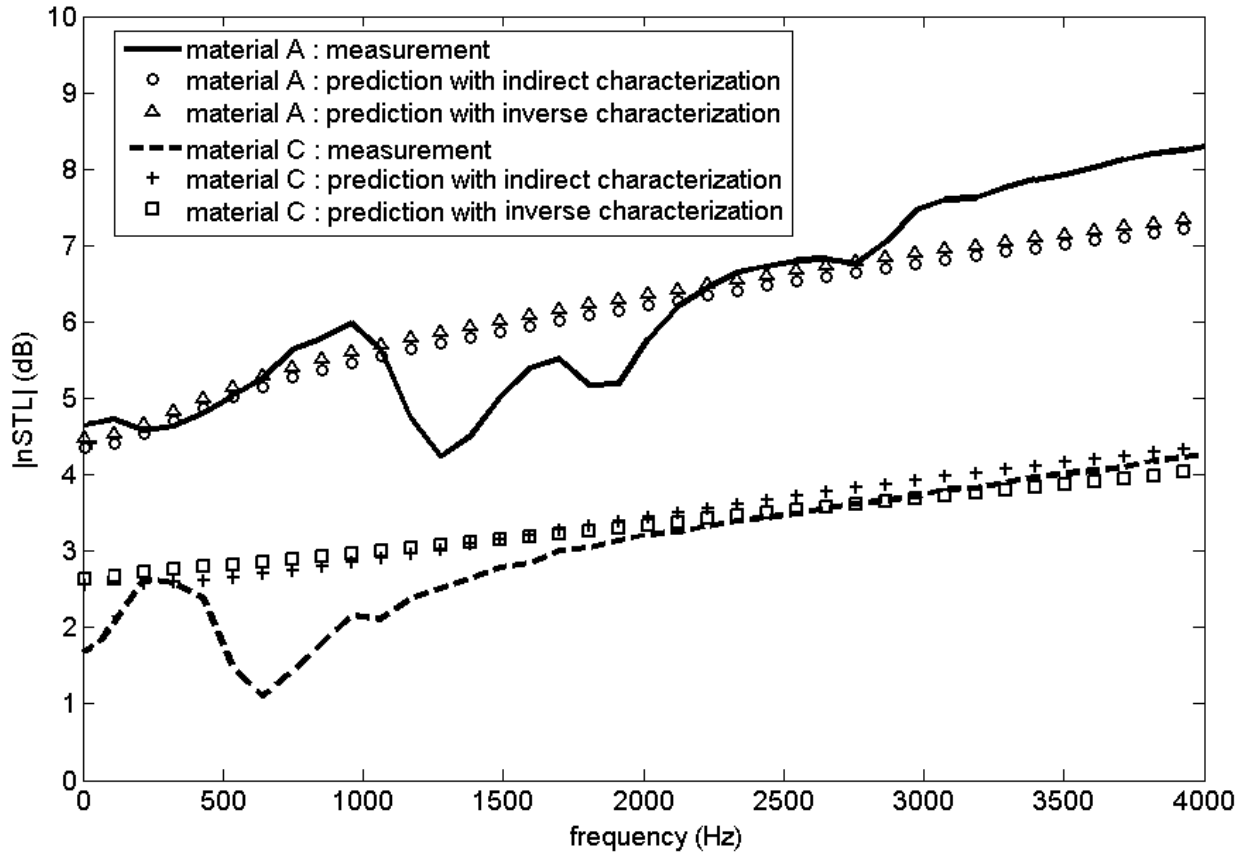

*Figure 3. Normal incidence sound transmission loss: comparison between measurements and predictions for materials A and C.*

## 5. Conclusion

In this paper, a straightforward application of an indirect method based on a three-microphone impedance tube setup to determine the non-acoustic properties of a sound absorbing porous material has been proposed and tested on four sound absorbing materials frequently encountered in the context of transport or building industries. This straightforward procedure only requires a direct measurement of the open porosity of the material and an impedance tube measurement where the sample is backed by a hard termination, as for classical sound absorption measurements (see Fig.1). An indirect characterization of the non-acoustic properties is performed and gives similar results compared to direct and inverse methods. Predictions of the normal incidence absorption coefficient and transmission loss using these macroscopic properties associated to the equivalent fluid model of Johnson-Champoux-Allard are in good agreement with impedance tube measurements.

**Acknowledgments**

The authors would like to thank the National Sciences and Engineering Research Council of Canada (NSERC) for providing financial support.**References**

1. J.F. Allard and N. Atalla, "Propagation of sound in porous media: Modeling sound absorbing materials", Second Edition, New York, *Elsevier Applied Science*, (2009).

2. Y. Salissou and R. Panneton, « Pressure/mass method to measure open porosity of porous solids, » *J. Applied Physics*, **101**, 124913.1-124913.7 (2007).

3. M.R. Stinson and G.A. Daigle, "Electronic system for the measurement of flow resistance," J. Acoust. Soc. Am., **83**, 2422-2428 (1988).

4. R.J.S. Brown, "Connection between formation factor for electrical resistivity and fluid-solid coupling factor in Biot's equations for acoustic waves in fluid-filled media," Geophys **45**, 1269-1275 (1980).

5. S. Brunauer, P.H. Emmett, and E. Teller, "Adsorption of gases in multimolecular layers," J. Am. Chem. Soc. **60**, 309-319 (1938).

6. R. Panneton, X. Olny, "Acoustical determination of the parameters governing viscous dissipation in porous media", J. Acoust. Soc. Am. **119**(4), 2027-2040, (2006).15